\documentstyle[11pt]{article}
\def\bea{\begin{eqnarray}}
\def\eea{\end{eqnarray}}

\def\beq{\begin{equation}}
\def\eeq{\end{equation}}
\def\ba{\beq\new\begin{array}{c}}
\def\ea{\end{array}\eeq}
\def\be{\ba}
\def\ee{\ea}

\def\2{{1\over 2}}

\makeatletter
\newdimen\normalarrayskip              
\newdimen\minarrayskip                 
\normalarrayskip\baselineskip
\minarrayskip\jot
\newif\ifold             \oldtrue            \def\new{\oldfalse}
\def\arraymode{\ifold\relax\else\displaystyle\fi} 
\def\eqnumphantom{\phantom{(\theequation)}}     
\def\@arrayskip{\ifold\baselineskip\z@\lineskip\z@
     \else
     \baselineskip\minarrayskip\lineskip2\minarrayskip\fi}
\def\@arrayclassz{\ifcase \@lastchclass \@acolampacol \or
\@ampacol \or \or \or \@addamp \or
   \@acolampacol \or \@firstampfalse \@acol \fi
\edef\@preamble{\@preamble
  \ifcase \@chnum
     \hfil$\relax\arraymode\@sharp$\hfil
     \or $\relax\arraymode\@sharp$\hfil
     \or \hfil$\relax\arraymode\@sharp$\fi}}
\def\@array[#1]#2{\setbox\@arstrutbox=\hbox{\vrule
     height\arraystretch \ht\strutbox
     depth\arraystretch \dp\strutbox
     width\z@}\@mkpream{#2}\edef\@preamble{\halign
\noexpand\@halignto
\bgroup \tabskip\z@ \@arstrut \@preamble \tabskip\z@ \cr}%
\let\@startpbox\@@startpbox \let\@endpbox\@@endpbox
  \if #1t\vtop \else \if#1b\vbox \else \vcenter \fi\fi
  \bgroup \let\par\relax
  \let\@sharp##\let\protect\relax
  \@arrayskip\@preamble}
\def\eqnarray{\stepcounter{equation}%
              \let\@currentlabel=\theequation
              \global\@eqnswtrue
              \global\@eqcnt\z@
              \tabskip\@centering
              \let\\=\@eqncr
              $$%
 \halign to \displaywidth\bgroup
    \eqnumphantom\@eqnsel\hskip\@centering
    $\displaystyle \tabskip\z@ {##}$%
    \global\@eqcnt\@ne \hskip 2\arraycolsep
         $\displaystyle\arraymode{##}$\hfil
    \global\@eqcnt\tw@ \hskip 2\arraycolsep
         $\displaystyle\tabskip\z@{##}$\hfil
         \tabskip\@centering
    &{##}\tabskip\z@\cr}
\begingroup\ifx\undefined\newsymbol \else\def\input#1 {\endgroup}\fi
\input amssym.def \relax
\input amssym
\newfont{\hr}{msbm10}
\newfont{\ams}{msam10}

\begin{document}
\setcounter{footnote}{1}
\def\thefootnote{\fnsymbol{footnote}}
\begin{center}
\hfill FIAN/TD-29/99,\ ITEP/TH-70/99\\
\hfill hep-th/9912088\\
\vspace{0.3in}
{\Large\bf Commuting Hamiltonians from Seiberg-Witten
$\Theta$-Functions}
\end{center}
\centerline{{\large A.Mironov}\footnote{Theory Dept.,
Lebedev Physical Inst. and ITEP, Moscow,
Russia}, {\large A.Morozov}\footnote{ITEP, Moscow,
Russia}}

\begin{quotation}
{Elementary MAPLE calculations are used to support the claim of
hep-th/9906240
that the ratios of theta-functions, associated with the
Seiberg-Witten complex curves,
provide Poisson-commuting Hamiltonians
which describe the dual of the original Seiberg-Witten integrable system.
}\end{quotation}

\bigskip

\setcounter{footnote}{0}

The goal of this note is to suggest a numerical evidence in favor of the
manifest construction proposed in \cite{dell} for the dual integrable systems
\cite{R,ITEP}. This construction allows one to build systems dual to the
elliptic Calogero and Ruijsenaars models with elliptic dependence on momenta
and ultimately leads to the (self-dual) double elliptic system describing
$6d$ SUSY gauge theories with the adjoint matter hypermultiplet. We start
with repeating and discussing the main steps of the construction \cite{dell}.

\noindent
{\bf 1.} Standard approach \cite{GKMMM} to
the Seiberg-Witten anzatz \cite{SW} is to
associate with every theory an integrable system ${\cal S}$ given by
a Lax operator $L(\vec p, \vec q|\xi)$ that naturally gives rise to
a family of the spectral curves
of genus $g$,
\be\label{sc}
{\cal C}: \ \ \ \det\left(L(\vec p, \vec q|\xi) - \lambda\right)
= 0
\ee
parametrized by the Poisson-commuting Hamiltonians (moduli)
$h_n(\vec p, \vec q)$, $n = 1, \ldots, g$,
\be
\left\{h_m, h_n\right\} \equiv
\frac{\partial h_m}{\partial \vec p}
\frac{\partial h_n}{\partial \vec q} -
\frac{\partial h_n}{\partial \vec p}
\frac{\partial h_m}{\partial \vec q} = 0
\ee
A distinguished set of the Poisson-commuting quantities is
provided by the ``flat'' moduli (action variables of the system ${\cal S}$)
\be
\vec a \equiv \oint_{\vec A} dS
\ee
which are the $A$-periods of the presymplectic 1-form $dS$
characterized by the property that its infinitesimal
variations with respect to moduli is holomorphic on the curve
${\cal C}$,
\be
\bar\partial \frac{\partial dS}{\partial\ {\rm moduli}} = 0
\ee
See refs.\cite{SWint,revs} for further details and references.

\noindent
{\bf 2.} In \cite{dell} it was claimed that another interesting set
of Poisson-commuting quantities is provided by the ratios of
Riemannian theta-functions (KP $\tau$-functions) associated with
the curve ${\cal C}$ -- and they were interpreted as the Hamiltonians
of the {\it dual} integrable system.

In the case of $GL(N)$, $N = g+1$ an explicit construction looks as
follows. The spectral curve of the original integrable system
(Toda chain, Calogero, Ruijsenaars or the most interesting double
elliptic system) has a period matrix $T_{ij}(\vec a)$,
$i,j = 1,\ldots,N$ with the special property:
\be
\sum_{j=1}^N T_{ij}(a) = \tau, \ \ \forall i
\label{sumT}
\ee
where $\tau$ does not depend on $a$.
As a corollary, the genus-$N$ theta-function is naturally decomposed
into a linear combination of genus-$g$ theta-functions:
\be
\Theta^{(N)}(p_i|T_{ij}) =
\sum_{n_i \in Z} \exp\left(i\pi\sum_{i,j=1}^NT_{ij}n_in_j +
2\pi i\sum_{i=1}^N n_ip_i\right) = \\
\sum_{k=0}^{N-1}
\theta_{\left[{k\over N},{0}\right]}(N\zeta|N\tau)
\cdot \check\Theta_k^{(g)}(\check p_i|\check T_{ij})
\ee
where
\be
\Theta_k \equiv
\check\Theta_k^{(g)}(\check p_i|\check T_{ij}) \equiv
\sum_{{n_i \in Z}\atop{\sum n_i = k}}
\exp\left(-i\pi\sum_{i,j=1}^N\check T_{ij}n_{ij}^2 +
2\pi i\sum_{i=1}^N n_i\check p_i\right)
\ee
and $p_i = \zeta + \check p_i$, $\sum_{i=1}^N \check p_i = 0$;
$T_{ij} = \check T_{ij} + \frac{\tau}{N}$,
$\sum_{j=1}^N \check T_{ij} = 0$, $\forall i$.

The claim of ref.\cite{dell} is that all the ratios
$\Theta_k/\Theta_l$ are Poisson-commuting with respect to the
Seiberg-Witten symplectic structure
\be
\sum_{i=1}^N dp_i\wedge da_i
\label{ss}
\ee
(where $a_N \equiv -a_1 - \ldots - a_{N-1}$, i.e.
$\sum_{i=1}^N a_i = 0$):
\be
\left\{\frac{\Theta_k}{\Theta_l}, \frac{\Theta_m}{\Theta_n}\right\} = 0
\ \ \forall k,l,m,n
\label{comm}
\ee
or
\be
\Theta_i\{\Theta_j, \Theta_k\} +
\Theta_j\{\Theta_k, \Theta_i\} +
\Theta_k\{\Theta_i, \Theta_j\} = 0  \ \ \forall i,j,k
\ee
or
\be
\{\log\Theta_i, \log\Theta_j\} =
\left\{\log\frac{\Theta_i}{\Theta_j}, \log\Theta_k\right\}
\ \ \forall i,j,k
\ee
The Hamiltonians of the dual integrable system can be chosen in
the form $H_k = \Theta_k/\Theta_0$, $k = 1,\ldots, g$.

\noindent
{\bf 3.} This claim was partly supported by the old observation
\cite{Kri,Kri2} that {\it zeroes} of the KP (Toda)
$\tau$-function (i.e. essentially the Riemannian theta-function), associated
with the spectral curve (\ref{sc}) are nothing but the coordinates $q_i$ of
the original (Calodero, Ruijsenaars) integrable system ${\cal S}$.
In more detail, due to the property (\ref{sumT}),
$\Theta^{(N)}(p|T)$ as a function
of $\zeta = \frac{1}{N} \sum_{i=1}^N p_i$ is an elliptic function on the
torus $(1,\tau)$ and, therefore, can be decomposed into an $N$-fold product
of the genus-one theta-functions. Remarkably, their arguments are just
$\zeta-q_i$:
\be
\Theta^{(N)}(p|T) =
c(p,T,\tau)\prod_{i=1}^N \theta(\zeta - q_i(p,T)|\tau)
\ee
(In the case of the Toda chain when $\tau\to
i\infty$ this "sum rule" is implied by the standard expression
for the individual $e^{q_i}$ through the KP $\tau$-function.) Since one
can prove that $q_i$ form a Poisson-commuting set of variables with respect to
the symplectic structure (\ref{ss}), this observation indirectly justifies
the claim of ref.\cite{dell}.

A deeper argument for the commutativity of such ratios
should come from the study of  (the quasiclassical limit of)
quantum $\tau$-functions and their properties \cite{qtau}
(i.e. from group theory),
but this is beyond the scope of the present letter.

Postponing discussion of deep theoretical origins of
(\ref{comm}), we report here some results of MAPLE calculations,
which provide a nontrivial check of those relations.

\noindent
{\bf 4.1.} In the simplest case of the Seiberg-Witten family
\be
w = P_N(\lambda) = \prod_{i=1}^N(\lambda - \lambda_i);
\ \ \ dS = \lambda d\log w
\label{pertTodacurve}
\ee
the flat moduli $a_i = \lambda_i$, the period matrix
is singular and only finite number of terms survives in the
series for the theta-function:
\be
\Theta^{(N)}(p|T) = \sum_{k=0}^{N-1} e^{2\pi ik\zeta}H^{(0)}_k(p,a),
\label{pertTodatheta}
\ee
\be
H^{(0)}_k(p,a) = \sum_{I, [I]=k} \prod_{i\in I}e^{2\pi ip_i}
\prod_{j\in \bar I} {\cal F}^{(0)}_{ij}(a)
\label{pertTodaHam}
\ee
Here
\be
{\cal F}^{(0)}_{ij}(a) = \frac{\Lambda}{a_{ij}}
\label{pertTodaF}
\ee
and $I$ are all possible partitions of $N$ indices into the sets
of $k = [I]$ and $N-k = [{\bar I}]$ elements. Parameter $\Lambda$
becomes significant only when the system is deformed:
either non-perturbatively or to more complex systems of the
Calogero--Ruijsenaars--double-elliptic  family.

This is the case of perturbative $4d$ pure $N=2$ SYM theory with
the prepotential
\be
{\cal F}^{(0)}(a) = \frac{1}{2i\pi}\sum_{i<j}^N
a_{ij}^2\log a_{ij}
\ee
The corresponding $\tau$-function $\Theta^{(N)}(p|T)$,
eq.(\ref{pertTodatheta}), describes an $N$-soliton solution
to the KP hierarchy. The Hamiltonians $H^{(0)}_k$,
eq.(\ref{pertTodaHam}), are those of the degenerated rational Ruijsenaars
system, and they are well-known to Poisson-commute with respect
to the relevant Seiberg-Witten symplectic structure (\ref{ss}).


The same construction for the other perturbative
Seiberg-Witten systems ends up with the Hamiltonians
of the more sophisticated Ruijsenaars systems.

For the spectral curves \cite{BMMM2}
\be
w={P_N(\lambda)\over P_N(\lambda-m)}={\prod_{i=1}^n(\lambda-\lambda_i)\over
\prod_{i=1}^n(\lambda-\lambda_i-m)};\ \ \ dS=\lambda d\log w
\ee
(perturbative $4d$ $N=4$ SYM with SUSY softly broken down
to $N=2$ by the mass $m$) the Poisson-commuting (with respect to the same
(\ref{ss})) Hamiltonians $H^{(0)}_k$ are given by
(\ref{pertTodaHam}) with
\be
{\cal F}^{(0)}_{ij}(a) = \frac{\sqrt{a_{ij}^2 - m^2}}{a_{ij}}
\label{pertCalF}
\ee
i.e. are the Hamiltonians of the rational Ruijsenaars system.

For the spectral curve \cite{BMMM2}
\be
w=e^{-2i\epsilon N}{P_N(\lambda)\over P_N(\lambda e^{-2i\epsilon})};
\ \ \ dS=\log\lambda d\log w
\ee
(perturbative $5d$ $N=2$ SYM compactified on a circle
with an $\epsilon$ twist as the boundary conditions)
the Hamiltonians are given by (\ref{pertTodaHam}) with
\be
{\cal F}^{(0)}_{ij}(a) =
\frac{\sqrt{\sinh(a_{ij}+\epsilon)\sinh(a_{ij}-\epsilon)}}
{\sinh a_{ij}}
\label{pertRuF}
\ee
i.e. are the Hamiltonians of the trigonometric Ruijsenaars system.

Finally, for the perturbative limit of the
most interesting self-dual double-elliptic system
\cite{dell} (the explicit form of its spectral curves is yet unknown)
the relevant Hamiltonians are those of the elliptic Ruijsenaars
system, given by the same (\ref{pertTodaHam}) with
\be
{\cal F}^{(0)}_{ij}(a) =\sqrt{1-{2g^2\over \hbox{sn}^2_{\tilde\tau}(a_{ij})}}
\sim\frac{\sqrt{\theta(\hat a_{ij}+\varepsilon|\tilde\tau)
\theta(\hat a_{ij}-\varepsilon|\tilde\tau)}}
{\theta(\hat a_{ij}|\tilde\tau)}
\label{pertdellF}
\ee
where $\tilde\tau$ is the modulus of the second torus associated with the
double elliptic system.

In all these examples $H_0^{(0)} = 1$, the theta-functions
$\Theta^{(N)}$ are singular and given by determinant (solitonic)
formulas with finite number of items (only terms with $n_i =0, 1$
survive in the series expansion of the theta function),
and Poisson-commutativity of arising Hamiltonians is analytically
checked within the theory of Ruijsenaars integrable systems.

\noindent
{\bf 4.2.} Beyond the perturbative limit, the analytical evaluation of
$\Theta^{(N)}$ becomes less straightforward.

The deformation of the curve (\ref{pertTodacurve}),
\be
w + \frac{\Lambda^{2N}}{w} = P_N(\lambda),
\ \ \ dS = \lambda d\log w
\ee
is associated with somewhat sophisticated prepotential
of the Toda chain integrable system,
\be
{\cal F}(a) = \frac{1}{2}\sum_{i<j}^N a_{ij}^2\log\frac{a_{ij}}{\Lambda}
+ \sum_{k=1}^\infty \Lambda^2{\cal F}^{(k)}(a)
\ee
The period matrix is
\be
T_{ij} = \frac{\partial^2{\cal F}}{\partial a_i\partial a_j} =
\tau\delta_{ij}\ + \log\frac{a_{ij}}{\Lambda} +
\sum_{k=1}^\infty \Lambda^2
\frac{\partial^2{\cal F}^{(k)}(a)}{\partial a_i\partial a_j},
\ \  i\neq j, \\ T_{ii} = \tau - \sum_{j\neq i} T_{ij}
\ee
and $\tau$ in this case can be removed by the rescaling of $\Lambda$.
Then,
\be
\Theta^{(N)}(p|T) = \sum_{k=0}^{N-1} e^{2\pi ik\zeta}
\Theta_k(p,a) = \\
= \sum_{k=0}^{N-1} e^{2\pi i k\zeta}
\sum_{n_i,\ \sum_i n_i = k}
e^{-i\pi\sum_{i<j} T_{ij}n_{ij}^2} e^{2\pi i \sum_i n_ip_i} = \\
= \left( 1 + \sum_{i\neq j}^N e^{2\pi i(p_i - p_j)}
{\cal F}_{ij}^4\prod_{k\neq i,j} {\cal F}_{ik}{\cal F}_{jk} + \right. \\
\left. +
\sum_{i\neq j\neq k\neq l}^N  e^{2\pi i(p_i + p_j - p_k - p_l)}
{\cal F}_{ik}^4 {\cal F}_{il}^4 {\cal F}_{jk}^4  {\cal F}_{jl}^4
\prod_{m\neq i,j,k,l} {\cal F}_{im}{\cal F}_{jm}{\cal F}_{km}{\cal F}_{lm}
+ \ldots\ \right) + \\
+ e^{2\pi i \zeta}
\left(\sum_{i=1}^N e^{2\pi i p_i} \prod_{j\neq i} {\cal F}_{ij} +
\sum_{i\neq j\neq k} e^{2\pi i (p_i + p_j - p_k)}
{\cal F}_{ik}^4{\cal F}_{jk}^4\prod_{l\neq i,j,k} {\cal F}_{il}{\cal F}_{jl}
{\cal F}_{kl} + \right.\\
\left. + \sum_{i\neq j} e^{2\pi i (2p_i - p_j)}
{\cal F}_{ij}^9\prod_{k\neq i,j} {\cal F}_{ik}^4 {\cal F}_{jk} + \ldots \
\right) + \\
+ e^{4\pi i\zeta}
\left(\sum_{i\neq j} e^{2\pi i(p_i+p_j)}\prod_{k\neq i,j}{\cal F}_{ik}
{\cal F}_{jk} +
\sum_i e^{4\pi ip_i} \prod_{k\neq i}{\cal F}_{ik}^4 + \ldots \ \right) + \\
+ \ldots
\label{Thetaexp}
\ee
with
\be
{\cal F}_{ij} = e^{-i\pi T_{ij}} = {\cal F}^{(0)}_{ij}
\left( 1 -
i\pi\frac{\partial^2{\cal F}^{(1)}(a)}{\partial a_i\partial a_j}
+ \ldots\ \right), \ \ \      i\neq j
\label{F}
\ee
The first few corrections  ${\cal F}^{(k)}$ to the prepotential
are explicitly known in the Toda-chain case \cite{IC},
for example,
\be
{\cal F}^{(1)} = -\frac{1}{2i\pi}\sum_{i=1}^N\prod_{k\neq i}
\left({\cal F}_{ik}^{(0)}\right)^2
\label{F1}
\ee

The coefficients $\Theta_k$ in (\ref{Thetaexp})
are expanded into powers of $(\Lambda/a)^{2N}$ and the leading (zeroth-order)
terms are exactly the perturbative expressions (\ref{pertTodaHam}).
Thus, the degenerated Ruijsenaars Hamiltonians (\ref{pertTodaHam}) are just the
perturbative approximations to the $H_k = \Theta_k/\Theta_0$ --
the full Hamiltonians of integrable system, dual to the Toda
chain\footnote{It would be interesting to compare this system with the
recently proposed "elliptic Toda" system \cite{Kri3}.}.

With the help of MAPLE we checked that the first corrections
to the degenerated Ruijsenaars Hamiltonians (\ref{pertTodaHam}) indeed
preserve Poisson-commutativity.  We did it up to the second order in
$\Lambda^{2N}$ for $N=3$ and up to the first order for $N=4$.

Note that there are three sources of deviations of
$H_k = \Theta_k/\Theta_0$ from $H^{(0)}_k$:

\begin{itemize}
\item the terms with some $n_i >1$ are taken into account
in the expansion (\ref{Thetaexp})

\item ${\cal F}_{ij} \neq  {\cal F}^{(0)}_{ij}$

\item $\Theta_0 \neq 1$
\end{itemize}

We checked that all these deviations are significant for the
Poisson-commutativity: it is important to use all the specifics
of the Seiberg-Witten
Riemannian theta-function to obtain the Poisson-commuting
Hamiltonians.

\noindent
{\bf 4.3.} Similar checks up to the first non-perturbative order
were performed for $N=3$ for the duals of

\begin{itemize}
\item
Calogero model -- i.e. with ${\cal F}_{ij}^{(0)}$ of the form
(\ref{pertCalF})

\item
Ruijsenaars model -- i.e. with ${\cal F}_{ij}^{(0)}$ of the form
(\ref{pertRuF})

\item
and the double-elliptic model of ref.\cite{dell} --
with ${\cal F}_{ij}^{(0)}$ of the form (\ref{pertdellF}); in this case
the check was made only with the first non-trivial correction in $\tilde
q=e^{2\pi i\tilde \tau}$
\end{itemize}

Note that the period matrix in these cases is expanded into powers of
$q=e^{2\pi i\tau}$:

\be
T_{ij} = \frac{\partial^2{\cal F}}{\partial a_i\partial a_j} =
\tau\delta_{ij}\ + \log\frac{a_{ij}}{\Lambda} +M^2
\sum_{k=1}^\infty q^k
\frac{\partial^2{\cal F}^{(k)}(a)}{\partial a_i\partial a_j},
\ \  i\neq j
\ee
and the dimensional constant $M$ depends on the system. E.g., in the Calogero
model $M=im$ etc. The limit to the Toda system corresponds to $q\to 0$, $q
M^{2N}$=fixed.

The result (Poisson-commutativity of $H_1 = \Theta_1/\Theta_0$
and $H_2 = \Theta_2/\Theta_0$ in this approximation)
significantly depends on the form (\ref{F1}) of the
first (instanton-gas)   correction to the prepotential.
It is indeed known to be of this form not only for the Toda
chain, but also  for Calogero system \cite{DP}. For
Ruijsenaars and double-elliptic systems eq.(\ref{F1})
is not yet available in the literature. We checked that
(\ref{F1}) is true for these two systems for $N=2$
(while we {\it used} (\ref{F1}) in calculations for $N=3$).

\noindent
{\bf 5.} In conclusion, we found new non-trivial evidence
in support of the claim \cite{dell} that the theta-functions
ratios $H_k = \Theta_k/\Theta_0$ in the case of Seiberg-Witten
integrable systems provide Poisson-commuting Hamiltonians
of the dual integrable systems. The real raison d'etre
(and a reasonable proof) of this property remains to be found.

Also, if the {\it universal} expressions like (\ref{F1})
in terms of perturbative ${\cal F}^{(0)}_{ij}$ -- the same   for all
the systems -- will be found for higher corrections
to the prepotentials\footnote{
To avoid possible confusion, the recurrent relation
\cite{LNS,GMMM,Mas} for the Toda-chain prepotential,
$$
\frac{\partial^2{\cal F}}{\partial\log\Lambda^2} \sim
\frac{\partial^2{\cal F}}{\partial\log\Lambda\partial a_i}
\frac{\partial^2{\cal F}}{\partial\log\Lambda\partial a_j}
\left.\frac{\partial^2}{\partial p_i\partial p_j}
\log\Theta_0\right|_{p=0}
$$
does {\it not} immediately provide such  {\it universal}
expressions. Already for ${\cal F}^{(1)}$ this relation gives
$$
{\cal F}^{(1)} \sim\sum_{i<j}^N \left({\cal F}^{(0)}_{ij}\right)^2
\prod_{k\neq i,j}^N  {\cal F}^{(0)}_{ik}{\cal F}^{(0)}_{jk}
$$
which coincides with (\ref{F1}) for the Toda-chain
${\cal F}^{(0)}_{ij}$, eq.(\ref{pertTodaF}), but is not true
(in variance with (\ref{F1})) for Calogero ${\cal F}^{(0)}_{ij}$,
eq.(\ref{pertCalF}). Meanwhile, the recurrent relations of ref.\cite{MNW} for
the Calogero system provide more promising expansion.

All this emphasizes once again the
need to study extension of the Whitham theory \cite{Whith,GMMM}
and WDVV-equations \cite{WDVV} from the Toda chains to the Calogero and
Ruijsenaars systems.
},  this will immediately give an explicit
(although not the most appealing) construction of the
Hamiltonians dual to the Calogero and Ruijsenaars models and -- especially
important -- the self-dual double-elliptic system which
was explicitly constructed in \cite{dell} only
for $N=2$.

We are indebted for useful discussions to H.W.Braden and A.Marshakov.
Our work is partly supported by
the RFBR grants 98-01-00328 (A.Mir.), 98-02-16575 (A.Mor.),
the Russian President's Grant 96-15-96939, the INTAS grant
97-0103 and the program for support of the scientific schools
96-15-96798. A.Mir. also acknowledges the Royal Society for
support under a joint project.

\end{document}